\documentclass[12pt]{article}
\usepackage{graphicx}
\usepackage{cite}

\newcommand{\be}{\begin{equation}}
\newcommand{\ee}{\end{equation}}
\newcommand{\bea}{\begin{eqnarray}}
\newcommand{\eea}{\end{eqnarray}}
\newcommand{\smallw}{{\scriptscriptstyle W}} %
\newcommand{\smallr}{{\scriptscriptstyle R}} %
\newcommand{\smalll}{{\scriptscriptstyle L}} %
\newcommand{\smallx}{{\scriptscriptstyle X}} %
\newcommand{\smallmax}{{\rm\scriptstyle max}} %
\newcommand{\smallmin}{{\rm\scriptstyle min}} %
\newcommand{\gl}{g_\smalll}
\newcommand{\gr}{g_\smallr}
\newcommand{\fl}{f_\smalll}
\newcommand{\fr}{f_\smallr}
\newcommand{\flr}{f_{\smalll \smallr}}
\newcommand{\fx}{f_\smallx}
\newcommand{\hov}{\bar{h}}
\newcommand{\flov}{\bar{f}_\smalll}
\newcommand{\frov}{\bar{f}_\smallr}
\newcommand{\flrov}{\bar{f}_{\smalll \smallr}}
\newcommand{\fxov}{\bar{f}_\smallx}

\newcommand{\mw}{M_\smallw}

\def \sss  {\scriptscriptstyle}

\def \gev  {\mbox{ GeV}}
\def \mev  {\mbox{ MeV}}
\def \dilog  {\mbox{Li}_2}

\def \psl  {p \kern-.45em{/}}
\def \qsl  {q \kern-.45em{/}}
\def \lsim {\raisebox{-.7ex}{$\stackrel{\textstyle <}{\sim}\,$}}

\def \Be7  {$\!\!\!\!\phantom{A}^7$Be~}
\def \B8   {$\!\!\!\!\phantom{A}^8$B~}
\def \bm   {\boldmath}
\def \eom  {E_\omega}
\def \zom  {z_\omega}
\def \notin {\in \!\!\!\!\!/}

\begin{document}              

%%%%%%%%%%%%%%%%%%%%%%%%%%%% titlepage %%%%%%%%%%%%%%%%%%%%%%%%%%%%%%%%%%%%
\begin{titlepage}
\begin{flushright}
        \small
        BUTP 2000/11\\
        November 2000
\end{flushright}

\renewcommand{\thefootnote}{\fnsymbol{footnote}}

\begin{center}
\vspace{2cm}
{\LARGE \bf QED Corrections to Neutrino \\Electron Scattering}

\vspace{1cm}
{\large\bf    M.~Passera\footnote
                {Email address: passera@itp.unibe.ch}}
\setcounter{footnote}{0}
\vspace{.5cm}

{\it    Institut f\"{u}r Theoretische Physik, 
        Universit\"{a}t Bern, \\
        Sidlerstrasse 5, CH-3012 Bern, Switzerland}
\vspace{1.8cm}

{\large\bf Abstract} 
\end{center} 
\vspace{5mm} 
\noindent 
We evaluate the $O(\alpha)$ QED corrections to the recoil electron
energy spectrum in the process $\nu_l + e \rightarrow \nu_l + e
\;(+\gamma)$, where $(+\gamma)$ indicates the possible emission of a
photon and $l=e$, $\mu$ or $\tau$. The soft and hard bremsstrahlung
differential cross sections are computed for an arbitrary value of the
photon energy threshold. We also study the $O(\alpha)$ QED corrections
to the differential cross section with respect to the total combined
energy of the recoil electron and a possible accompanying photon.
Their difference from the corrections to the electron spectrum is
investigated. We discuss the relevance and applicability of both
radiative corrections, emphasizing their role in the analysis of
precise solar neutrino electron scattering experiments.

\end{titlepage}

%%%%%%%%%%%%%%%%%%%%%%%%%%%%%%%%%%%%%%%%%%%%%%%%%%%%%%%%%%%%%%%%%%%%%%%%%%%%%
\section{Introduction}

The QED corrections to the scattering process $\nu_{e} + e \rightarrow
\nu_{e} + e$ were studied long ago by Lee and Sirlin using the
effective four--fermion V$-$A Lagrangian \cite{LS}, and shortly
afterwards Ram \cite{Ram} extended their calculations by including
hard photon emission. A few years later, 't Hooft computed the lowest
order prediction to this differential cross section in the Standard
Model (SM) \cite{tH}. Since then, the radiative corrections to this
process, which plays a fundamental role in the study of electroweak
interactions, have been investigated by many authors, focusing on
various aspects of the problem
\cite{ZKN,SU,GV,MS,AHKKM,AH,MSS,DB1,DB2,BBBS,Be2000}.

't Hooft's early SM predictions for $\nu$--$e$ scattering were used by
Bahcall to examine the total cross section, energy spectrum and
angular distribution of recoil electrons resulting from the scattering
with solar neutrinos \cite{B87}. More recently, Bahcall, Kamionkowski
and Sirlin performed a detailed investigation of the radiative
corrections to these recoil electron spectra and total cross sections
\cite{BKS}. Their results show the importance of these corrections for
the analysis of precise solar $\nu$--$e$ scattering experiments,
particularly of those measuring the higher energy neutrinos that
originate from \B8 decay.

In this paper we study the $O(\alpha)$ QED corrections to neutrino
electron scattering in the SM, with contributions involving either
neutral currents (as in the $\nu_{\mu,\tau} + e \rightarrow
\nu_{\mu,\tau} + e$ process) or a combination of neutral and charged
currents (as in the $\nu_{e} + e \rightarrow \nu_{e} + e$ process).
In this analysis we make the approximation of neglecting terms of
$O(q^2/\mw^2)$, where $q^2$ is the squared four-momentum transfer and
$\mw$ is the $W$ boson mass.  Within this approximation, which is
excellent for present experiments ($|q^2/\mw^2| \!\sim\! 1$ when the
electron recoil energy $\sim 6 \times 10^3$ TeV!), the SM radiative
corrections to these processes can be naturally divided into two
classes. The first, which we will call ``QED'' corrections, consist of
the photonic radiative corrections that would occur if the theory were
a local four--fermion Fermi theory rather than a gauge theory mediated
by vector bosons; the second, which we will refer to as the
``electroweak'' (EW) corrections, will be the remainder. The EW
corrections have been studied by several authors
\cite{SU,GV,MS,AHKKM,MSS,BKS} and are not discussed in the present
paper. The split--up of the QED corrections is sensible as they form a
finite (both infrared and ultraviolet) and gauge--independent subset
of diagrams. We refer the reader to ref.~\cite{Si78-80} for a detailed
study of this separation.

The QED radiative corrections are due to both loop diagrams (virtual
corrections) and to the bremsstrahlung radiation (real photons)
accompanying the scattering process. Of course, only this combination
of virtual and real photon corrections is free from infrared
divergences. To order $\alpha$, the bremsstrahlung events correspond
to the inelastic process $\nu_l + e \rightarrow $ $\nu_l + e +\gamma $
($l=e,\mu$ or $\tau$). Experimentally, bremsstrahlung events in which
photons are too soft to be detected are counted as contributions to
the elastic scattering $\nu_l + e \rightarrow$ $\nu_l + e$. The cross
section for these events should be therefore added to the theoretical
prediction of the elastic cross section, thus removing its infrared
divergence.

We will divide the bremsstrahlung events into ``soft'' (hereafter SB)
and ``hard'' (hereafter HB), according to the energy of the photon
being respectively lower or higher than some specified threshold
$\epsilon$. We should warn the reader that the words ``soft'' and
``hard'' may be slightly deceiving. Indeed, if $\epsilon$ is large
(small), the SB (HB) cross section will also include events with
relatively high (low) energy photons.  While calculations of both soft
and hard bremsstrahlung are often performed under the assumption that
$\epsilon$ is a very small parameter, much smaller than the mass of
the electron or its final momentum, we will also discuss
results for the case in which $\epsilon$ is an arbitrary parameter
constrained only by the kinematics of the process.  Indeed, the HB
cross section (contrary to the SB one) is by itself, at least in
principle, a physically measurable quantity for any kinematically
allowed value of this threshold. All calculations have been carried
out without neglecting the electron mass.

The paper is organized as follows. In sect.~2 we discuss the lowest
order prediction for the neutrino electron differential cross
section, together with its soft bremsstrahlung and one--loop QED
corrections. The hard bremsstrahlung recoil electron spectrum is
examined in sect.~3. In the same section, this contribution is added
to the virtual and soft ones to derive the total QED corrections.  In
sect.~4 we evaluate the spectrum of the total combined energy of the
recoil electron and a possible accompanying photon emitted in the
scattering process. We summarize our main results in sect.~5,
discussing their applicability and emphasizing their role in the
analysis of solar neutrino electron scattering experiments.

%2222222222222222222222222222222222222222222222222222222222222222222222
\section{Virtual and Soft Photon Corrections}

The SM prediction for the elastic neutrino electron differential
cross section is, in lowest--order and neglecting terms of
$O(q^2/\mw^2)$ \cite{tH},
\be
   \left[\frac{d\sigma}{dE}\right]_0 \;=\; \frac{2mG_{\mu}^2}{\pi}
        \left[\gl^2 +\gr^2 \left(1-z\right)^2 -\gl \gr 
        \left(\frac{m z}{\nu}\right)\right],
\label{eq:treelevel}
\ee
where $m$ is the electron mass, $G_{\mu}=1.16637(1) \times
10^{-5}\gev^{-2}$ is the Fermi coupling constant \cite{FOS}, $\gl =
\sin^2 \!\theta_{\smallw} \pm 1/2$ (upper sign for $\nu_e$, lower sign
for $\nu_{\mu,\tau}$), $\gr = \sin^2 \!\theta_{\smallw}$ and
$\sin\theta_{\smallw}$ is the sine of the weak mixing angle. In this
elastic process $E$, the electron recoil energy, ranges from $m$ to
$E_\smallmax =$ $[m^2 +(2\nu +m)^2]/[2(2\nu +m)]$, $z=(E-m)/\nu$ and
$\nu$ is the incident neutrino energy in the frame of reference in
which the electron is initially at rest.  We will refer to the $L$,
$R$ and $LR$ parts of an expression to indicate its terms proportional
to $\gl^2$, $\gr^2$ and $\gl \gr$, respectively. For example, the $R$
part of $\left[d\sigma/dE\right]_0$ (eq.~(\ref{eq:treelevel})) is
$(2mG_{\mu}^2/\pi) \gr^2 (1-z)^2$.

According to the definition discussed earlier, the one--loop QED
corrections to neutrino electron scattering consist of the photonic
vertex corrections (together with the diagrams involving the field
renormalization of the electrons) computed with the local
four--fermion Fermi Lagrangian.  These corrections give rise to the
following expression for the differential cross section:
\be 
   \left[\frac{d\sigma}{dE}\right]_{\rm Virtual}
   \;=\; \frac{2mG_{\mu}^2}{\pi} 
   \Biggl[\frac{\alpha}{\pi} \,\delta(E,\nu)\Biggr],
\label{eq:V}
\ee
where 
\bea \delta(E,\nu) &=& \gl^2\; \left\{V_1(E) +V_2(E)\left[
    z-1-\frac{mz}{2\nu} \right]
\right\} \nonumber  \\
&+& \gr^2\; \left\{V_1(E)\left(1-z\right)^2 + V_2(E)\left[
    z-1-\frac{mz}{2\nu} \right]
\right\} \nonumber  \\
&-& \gl\gr\,\left\{ \left[V_1(E)-V_2(E)\right] \left(\frac{mz}
    {\nu}\right) +2V_2(E) \left[z-1-z^2 \right] \right\}, 
\eea
\bea V_1(E) &=& \left(2\ln\!\frac{m}{\lambda}\right)
\left[1-\frac{E}{2l} \ln\!\left(\frac{E+l}{E-l}\right) \right] -2
-\frac{E}{l}\left[ \,\dilog\!\left(\frac{l-E+m}{2l}\right)
\right.     \nonumber  \\
&-& \left. \dilog\!\left(\frac{l+E-m}{2l}\right) \right] +
\frac{1}{4l} \left[3E+m-E\ln\!\left(\frac{2E+2m}{m}\right)\right]
\ln\!\left(\frac{E+l}{E-l}\right), \\ ~ \nonumber\\
V_2(E) &=& \frac{m}{4l}\ln\!\left(\frac{E+l}{E-l}\right).  
\eea
$\lambda$ is a small photon mass introduced to regularize the infrared
divergence and $l=\sqrt{E^2-m^2}$ is the three-momentum of the
electron. The dilogarithm $\dilog(x)$ is defined by
$$ \dilog(x) = -\int_0^x \!dt \,\frac{\ln(1-t)}{t}.  $$
The $L$ part of eq.~(\ref{eq:V}) (with $\gl=1$) is identical to the
formula for the one--loop photonic corrections to the $\nu_e +e
\rightarrow \nu_e +e$ differential cross section computed long ago in
the pioneering work of Lee and Sirlin \cite{LS} using the
effective four--fermion Fermi V$-$A Lagrangian. The analogous formula
for the reaction involving an anti--neutrino $\overline{\nu}_e$
(rather than a neutrino $\nu_e$) can be found in the same article and
coincides\footnote{With the exception of a minor typographical error
  in their eq.~22, where the square bracket multiplying 
  $I_{\rm rad}$ should be
  squared. We thank Alberto Sirlin for confirming this point.}  with
the $R$ part of eq.~(\ref{eq:V}) (with $\gr = 1$). This identity is
simply due to the fact that the cross section for antineutrinos in the
local V$-$A theory is the same as that for neutrinos calculated with a
V$+$A coupling. On the contrary, the $LR$ part of eq.~(\ref{eq:V}) has
clearly no analogue in the V$\pm$A theory, but can be derived very easily 
once the $L$ and $R$ parts are known.

The $\nu_e +e\rightarrow$ $\nu_e +e +\gamma$ differential cross
section with emission of a soft photon was computed in ref.~\cite{LS},
once again by using the effective four--fermion Fermi V$-$A
Lagrangian. It can be identified with the $L$ part (with $\gl=1$)
of the soft photon corrections to the tree level result in 
eq.~(\ref{eq:treelevel}). The $L$, $R$ and $LR$ parts of these corrections 
(with $\gl=\gr=1$) are however identical,
because the whole soft bremsstrahlung cross section is proportional to
its lowest--order elastic prediction.  We can therefore write the soft
photon emission cross section in the following factorized form:
\be
   \left[\frac{d\sigma}{dE}\right]_{\rm SB}
   \;=\; \frac{\alpha}{\pi} \;I_\gamma (E,\epsilon)
   \left[\frac{d\sigma}{dE}\right]_0,
\label{eq:SB}
\ee
with 
\bea
    I_\gamma(E,\epsilon) &=& \left(2\ln\!\frac{\lambda}{\epsilon}\right)
    \left[1-\frac{E}{2l} \ln\!\left(\frac{E+l}{E-l}\right) \right]
    + \frac{E}{2l}\left\{\,
    L\! \left(\frac{E+l}{E-l}\right) -L\! \left(\frac{E-l}{E+l}\right)
                  \right. \nonumber  \\
    &+& \left.    \ln\!\left(\frac{E+l}{E-l}\right) 
        \left[1-2\ln\!\left(\frac{l}{m}\right)\right] \right\} +1-2\ln\!2
\eea
and
$$
   L(x) = \int_0^x \!dt \,\frac{\ln|1-t|}{t}.  
$$ 
(For $x\in \mbox{I}\!\mbox{R}$, $L(x) = -\mbox{Re} [\dilog(x)]$.)
This result is valid under the assumption that $\epsilon$, the maximum
soft photon energy, is much smaller than $m$ or the final momentum of
the electron.  As we mentioned earlier, in the next section we will
discuss numerical results for the case in which $\epsilon$ is an
arbitrary parameter.

The reader will notice that the sum $[V_1(E)+I_\gamma(E,\epsilon)]$
does not depend on $\lambda$, the infrared regulator. Indeed, the
infrared divergence of the virtual corrections (eq.~(\ref{eq:V})) is
canceled by that arising from the soft photon emission
(eq.~(\ref{eq:SB})).

%3333333333333333333333333333333333333333333333333333333333333333333333
\section{Hard Bremsstrahlung and Total QED Corrections to the 
Final Electron Spectrum}

The SM prediction for the differential neutrino electron cross
section 
\be
     \nu_l + e \rightarrow \nu_l + e \;(+\gamma), 
\label{eq:nuegamma}
\ee
where $(+\gamma)$ indicates the possible emission of a photon, can be
cast, up to corrections of $O(\alpha)$, in the following form:
\bea
   \left[\frac{d\sigma}{dE}\right]_{\rm SM}
   & = & \frac{2mG_{\mu}^2}{\pi} 
        \Biggl\{\gl^2(E) \left[1+\frac{\alpha}{\pi} \fl(E,\nu) \right]
        +\gr^2(E) \left(1-z\right)^2 
        \left[1+\frac{\alpha}{\pi} \fr(E,\nu) \right] \nonumber\\
   & &  -\gl(E) \gr(E) \left(\frac{m z}{\nu}\right)
        \left[1+\frac{\alpha}{\pi} \flr(E,\nu) \right] \Biggr\}.
\label{eq:SMdE}
\eea 
(We remind the reader that terms of $O(q^2/\mw^2)$ are neglected
throughout this paper.)  The deviations of the functions $\gl(E)$ and
$\gr(E)$ from the lowest--order values $\gl$ and $\gr$ reflect the
effect of the electroweak corrections, which have been studied by
several authors \cite{SU,GV,MS,AHKKM,MSS,BKS}. (See ref.~\cite{BKS}
for simple numerical results.)

The functions $\fx(E,\nu)$ ($X=L,R$ or $LR$) describe the QED effects
(real and virtual photons). For simplicity of notation their $\nu$
dependence will be dropped in the following. Each of these functions
is the sum of virtual (V), soft (SB) and hard (HB) corrections,
\be
    \fx(E) = \fx^{\sss V}(E) + 
             \fx^{\sss SB}(E,\epsilon) + 
             \fx^{\sss HB}(E,\epsilon).
\label{eq:fxsum}
\ee
(We remind the reader that we have defined the bremsstrahlung events
as ``soft'' or ``hard'' according to the energy of the photon being
respectively smaller or higher than a specified threshold
$\epsilon$.) The analytic expressions for $\fx^{\sss V}(E)$ and
$\fx^{\sss SB}(E,\epsilon)$ can be immediately read from
eqs.~(\ref{eq:V}) and (\ref{eq:SB}) respectively (the latter being
valid only in the small $\epsilon$ limit) and their sums, which are
infrared--finite, will be denoted by
\be
    \fx^{\sss VS}(E,\epsilon) = \fx^{\sss V}(E) +
                          \fx^{\sss SB}(E,\epsilon).
\ee
Analytic expressions from which one can obtain $\fl^{\sss
  HB}(E,\epsilon)$ and $\fr^{\sss HB}(E,\epsilon)$ were calculated
long ago by Ram \cite{Ram}. Although these results were obtained in
the small $\epsilon$ approximation, the formulae are nonetheless long and
complicated and we will only plot the results for specific values
of $\nu$ and $\epsilon$. The function $\flr^{\sss HB}(E,\epsilon)$
has not been previously calculated.

We created {\tt BC}, a combined {\tt Mathematica}--{\tt FORTRAN}
code\footnote{The code {\tt BC}, available upon request, computes all
  QED corrections discussed in this paper. It uses the {\tt
    Mathematica} package {\tt FeynCalc} \cite{FC} and the {\tt
    FORTRAN} code {\tt VEGAS} \cite{VE}.} to compute the $\fx^{\sss
  HB}(E,\epsilon)$ functions for arbitrary 
positive values of the parameter
$\epsilon$ up to the kinematic limit $\nu$ ($\nu$, the incident
neutrino energy in the laboratory system, is also the maximum possible
energy of the emitted photon). We first computed the transition
probability for the bremsstrahlung process, averaged over the initial
electron spins and summed over the polarizations of the final electron
and photon. We then removed the energy--momentum conserving
$\delta$ function in the three--body phase--space integral and
specialized the result to the laboratory frame of reference. For a
given initial neutrino energy, the five independent variables
describing the final state were chosen to be the four angular
variables of the final electron and photon, plus the electron recoil
energy. In order to compute the HB functions $\fx^{\sss
  HB}(E,\epsilon)$ we finally imposed the condition that the photon
energy $\omega$ should be larger than the threshold $\epsilon$
(without assuming $\epsilon$ to be small). This constraint required a
detailed analysis of the kinematically allowed ranges of variability of
the chosen phase--space variables. The last integrations over
the angular variables were then performed numerically using the Monte
Carlo method \cite{VE} and demanding a $0.1\%$ relative accuracy.
(Reminding the reader that $X$ indicates $L$, $R$ or $LR$, we
note that this uncertainty in the computation of a function $\fx^{\sss
HB}(E,\epsilon)$ produces an extremely tiny relative error
$(\alpha/\pi)\fx^{\sss HB}(E,\epsilon)\times 0.1\%$ in the
corresponding $X$ part of the differential cross section in
eq.~(\ref{eq:SMdE}). This high level of accuracy was useful for
internal numerical checks and for the comparisons, in the small
$\epsilon$ limit, with Ram's results.)  All calculations have been
carried out without neglecting the electron mass.

In figs.~1 and 2 the functions $\fx^{\sss HB}(E,\epsilon)$ are
respectively plotted for $\nu = 0.862\mev$ and $\nu = 10\mev$, setting
the threshold $\epsilon$ to several different values.  These two values
of $\nu$ were chosen for their relevance in the study of solar
neutrinos: $\nu=0.862\mev$ is the energy of the monochromatic
neutrinos produced by electron capture on \Be7 in the solar interior,
while $\nu = 10\mev$ belongs to the continuous energy spectrum of the
solar neutrinos that originate from \B8 decay. In the same figures we
compared the results of our code {\tt BC} (solid lines) with the
approximate analytic results of Ram (dotted lines). (There are no
dotted lines in the $LR$ plots because the function $\flr^{\sss
  HB}(E,\epsilon)$ has not been previously calculated.) As we
mentioned earlier, Ram's formulae were computed in the small
$\epsilon$ approximation, keeping the logarithmically divergent terms
proportional to $\ln(\epsilon/m)$, but neglecting the remaining
$\epsilon$--dependent terms. Our results confirm Ram's ones in the
small $\epsilon$ limit. If $\epsilon$ is not small, the discrepancy
between solid and dotted curves increases with increasing values of
$\epsilon$. Figures 1 and 2 also show that the dotted curves are not
always positive. This is of course an unphysical property because the
HB differential cross section, being a transition probability for a
physical process, cannot be negative. (We also note that, in full
generality, we could set $\gl$ or $\gr$ to zero, in which case the HB
differential cross section would consist only of its $R$ or $L$ parts,
respectively.)  Our functions $\fl^{\sss HB}(E,\epsilon)$ and
$\fr^{\sss HB}(E,\epsilon)$ are always positive (or zero).

The total $O(\alpha)$ QED corrections $\fl(E)$ and $\fr(E)$, given by
the sum of V, SB and HB contributions (see eq.~(\ref{eq:fxsum})), can
be easily obtained by adding the analytic results of eqs.~(\ref{eq:V})
and (\ref{eq:SB}) to Ram's HB (lengthy) ones. Both SB and HB
corrections were computed in the small $\epsilon$ approximation, and
the logarithmically divergent terms proportional to $\ln(\epsilon/m)$
exactly drop out upon adding these soft and hard contributions. The
remaining $\epsilon$--dependent terms, which were neglected in both SB
and HB calculations, must cancel in the sum as well, and are therefore
irrelevant in the computation of the total QED corrections of
eq.~(\ref{eq:SMdE}). The $LR$ case is slightly different: Ram's
formulae, which were used to derive the small $\epsilon$ approximation
for $\fl^{\sss HB}(E,\epsilon)$ and $\fr^{\sss HB}(E,\epsilon)$, do
not provide us with the corresponding $LR$ correction. In order to
compute $\flr(E)$ we have therefore added the V and SB analytic
results of eqs.~(\ref{eq:V}) and (\ref{eq:SB}) to our HB numerical
results.  The ``exact'' $\epsilon$ dependence of our HB results is not
completely canceled by that of the SB, which includes only terms
proportional to $\ln(\epsilon/m)$, and the sum $\flr(E)$ contains therefore 
a residual (not logarithmically divergent) dependence on the photon
energy threshold $\epsilon$. This spurious dependence has been minimized by
fixing $\epsilon$ to be a very small value $\epsilon_{\smalll
  \smallr}$ chosen so as to have an estimated induced relative error
as small as $O(0.1\%)$. \footnote{With the exception of $E$ belonging
  to a tiny interval of $O(\epsilon_{\smalll \smallr})$ at the
  endpoint $E_\smallmax$.  We remind the reader that $0.1\%$ is also
  the relative numerical uncertainty used by our code {\tt BC} in the
  computation of the functions $\fx^{\sss HB}(E,\epsilon)$ and
  produces a totally negligible relative error $(\alpha/\pi)\fx^{\sss
    HB}(E,\epsilon)\times 0.1\%$ in the corresponding $X$ parts of the 
  differential cross section in eq.~(\ref{eq:SMdE}).}

Equations (\ref{eq:V}) and (\ref{eq:SB}) determine the analytic expression
of $\fx^{\sss VS}(E,\epsilon)$ (the infrared--finite sum of V and SB
corrections) in the small $\epsilon$ approximation. But the complete
$\epsilon$ dependence of our numerical $\fx^{\sss HB}(E,\epsilon)$
computations, combined with the knowledge of the above described
$\fx(E)$ functions, allows us to determine also the ``exact''
$\fx^{\sss VS}(E,\epsilon)$ functions via the subtraction
\be
\fx^{\sss VS}(E,\epsilon) =\fx(E)-\fx^{\sss HB}(E,\epsilon).
\label{eq:vsexact}
\ee
These will be the ``exact'' VS corrections employed in the rest of our
analysis.

In fig.~3 we plotted the functions $\fx(E)$ (thick solid), $\fx^{\sss
  HB}(E,\epsilon)$ (medium solid) and $\fx^{\sss VS}(E,\epsilon)$
(thin solid) for $\nu=0.862 \mev$. The threshold $\epsilon$ in the VS
and HB functions was set to $0.02\mev$. In figs.~4 and 5 we plotted
the same functions with $\nu=10 \mev$ ($\epsilon=1\mev$) and
$\nu=1\gev$ ($\epsilon=50\mev$), respectively.

In figs.~3, 4 and 5 we also plotted the simple approximate formulae
for $\fx(E)$ introduced in ref.~\cite{BKS} (dotted lines). These compact
analytic expressions were obtained by modifying the expressions of
ref.~\cite{MSS}, which had been evaluated in the extreme relativistic
approximation. (The $LR$ term of the differential cross section, being
proportional to $(m/\nu)$, vanishes in the extreme relativistic limit
and, therefore, cannot be derived from ref.~\cite{MSS}. As a consequence,
the $LR$ approximation of ref.~\cite{BKS} is only a (very educated!)
guess.)  Thanks to their simplicity, the compact formulae of
ref.~\cite{BKS} are easy to use and are employed, for example, by the
Super--Kamiokande collaboration \cite{SK} in their Monte Carlo simulations
for the analysis of the solar neutrino energy spectrum.

As it was noted in refs.~\cite{Ram, BKS}, all $\fx(E)$ functions contain a
term which diverges logarithmically at the end of the spectrum. This
feature, related to the infrared divergence, is similar to the one
encountered in the QED corrections to the $\mu$--decay spectrum
\cite{BFS,KS}. If $E$ gets very close to the endpoint we have
$(\alpha/\pi)\fx(E) \approx -1$, clearly indicating a breakdown of the
perturbative expansion and the need to consider multiple-photon
emission. However, this divergence
can be easily removed (in agreement with the KLN theorem
\cite{KS,KLN}) by integrating the differential cross section over
small energy intervals corresponding to the experimental energy
resolution. We also note that the singularity of
$\flr(E)$ for $E=m$ does not pose a problem, as the product
$(mz/\nu)\flr(E)$, which appears in the $LR$ part of the differential
cross section, is finite in the limit $E\rightarrow m$.  This can be
seen from the dashed line in the $LR$ plot of fig.~3, which indicates 
the function $(mz/\nu)\flr(E)$. In the same plot, the dot-dashed line
is the product of the $\flr(E)$ approximation of ref.~\cite{BKS} and 
$(mz/\nu)$.

%4444444444444444444444444444444444444444444444444444444444444444444444
\section{Spectrum of the Combined Energy of Electron and Photon}

We will now turn our attention to the analysis of the differential
$\nu_l + e \rightarrow$ $\nu_l + e \;(+\gamma)$ cross section relevant
to experiments measuring the {\em total combined energy} of the recoil
electron and a possible accompanying photon emitted in the scattering
process. We will begin by considering bremsstrahlung events with a
photon of energy $\omega$ larger than the usual threshold $\epsilon$
(HB).

The HB differential cross section $[d\sigma/d(E+\omega)]_{\rm HB}$ can
be immediately derived from the HB corrections to the energy spectrum
of the final neutrino.  In the elastic reaction $\nu_l + e
\rightarrow$ $\nu_l + e$, the final neutrino energy $\nu'$ ranges from
$\nu'_\smallmin=$ $\nu m/(2\nu+m)$ to $\nu'_\smallmax=\nu$ (the value
$\nu'=$ $\nu'_\smallmin$ occurs when the final electron and neutrino
are scattered back to back, with the electron moving in the forward
direction with $E=E\smallmax$; the value $\nu'=$ $\nu'_\smallmax$
occurs in the forward scattering situation).  When a photon of energy
$\omega > \epsilon$ is emitted, $\nu'$ varies between $0$ and 
$\nu-\epsilon$. If we now define the HB functions
$$
        h(\nu',\epsilon,\nu) \equiv \left[ \frac{d\sigma}{d\nu'} 
        \right]_{\rm HB}
        \qquad \mbox{and} \qquad
        \hov(E+\omega,\epsilon,\nu) \equiv 
        \left[\frac{d\sigma}{d(E+\omega)} \right]_{\rm HB}
$$
such that 
$$
    \int_0^{\nu-\epsilon} h(\nu',\epsilon,\nu) \,d\nu' \;=\; 
    \int_{m+\epsilon}^{\nu+m} \hov(E+\omega,\epsilon,\nu)\, d(E+\omega),
$$
conservation of energy implies then
$$
         \hov(E+\omega,\epsilon,\nu) = h(\nu',\epsilon,\nu) = 
         h(\nu+m-(E+\omega),\epsilon,\nu),
$$
and the HB differential cross section with respect to the sum of the
electron and photon energies can be directly obtained from 
$[d\sigma/d\nu']_{\rm HB}$. The variable $E+\omega$ varies between
$m+\epsilon$ and $m+\nu$ (note that $m+\nu = E_\smallmax + 
\nu'_\smallmin$).

The function $[d\sigma/d\nu']_{\rm HB}$, computed by our code {\tt
  BC}, has been evaluated in a manner similar to the HB corrections to
the electron recoil energy spectrum (see sect.~3). For a given initial
neutrino energy, the five independent variables of the three--body
phase--space describing the final state have been chosen to be the
four angular variables of the final neutrino and photon, plus the
final neutrino energy. In analogy with the case of the electron
spectrum, we imposed the condition $\omega>\epsilon$ (once again, the
threshold $\epsilon$ is not assumed to be small and can vary up to the
kinematic limit $\nu$) and evaluated the corresponding bounds on the
chosen kinematic variables. Just as for the $\fx^{\sss
  HB}(E,\epsilon)$ functions of sect.~3, the last integrations over
the angular variables were performed numerically using the Monte Carlo
method and requiring a very high (0.1\%) relative accuracy.

A check of the consistency of our results was performed by comparing
the values of the total HB cross section $\sigma_{\rm
  HB}(\nu,\epsilon)$ obtained by integrating the differential HB cross
sections of sects.~3 and 4. The equality
$$
        \sigma_{\rm HB}(\nu,\epsilon)   \;\;=\;\; 
                \int_m^{E_{\smallmax}} 
                \left[\frac{d\sigma}{dE}\right]_{\rm HB}
                \!\!\! dE                               \;\;=\;\;       
                \int_{m+\epsilon}^{\nu+m}
                \left[\frac{d\sigma}{d(E+\omega)}\right]_{\rm HB}
                \!\!\! d(E+\omega)
$$
has been tested for several values of $\nu$ and $\epsilon$, and all
relative deviations were found to be smaller than 0.1\% (which is also
the relative accuracy of the integrands).

We now combine virtual, soft and hard bremsstrahlung contributions in
order to evaluate the complete $O(\alpha)$ QED prediction for the
differential cross section $d\sigma/d(E+\omega)$ of reaction
(\ref{eq:nuegamma}). In sect.~3 we computed the total QED corrections
by simply adding these three parts. Their sum does not depend on
the threshold $\epsilon$. The combination of VS and HB terms requires
here a more careful analysis. We proceeded as follows. 
Let's consider an experimental setup for $\nu$--$e$ scattering able to
measure the photon energy if it's higher than a threshold $\epsilon$,
but completely blind to low energy photons $(\omega\!  <\!\epsilon)$.
Let's also assume that the electron energy $E$ is precisely measurable
independently of its value.  This detector can therefore measure the
usual electron spectrum $d\sigma/dE$ as well as the differential cross 
section $d\sigma/dE_\omega$, where the variable $\eom$ is
defined as follows,
\be 
    E_\omega \equiv \left\{
    \begin{array}{ll}
    E+\omega & \quad\mbox{if} \;\; \omega \geq \epsilon, \\
    E & \quad\mbox{if} \;\; \omega < \epsilon.
                \end{array} \right.
\label{eq:eom}
\ee
Figure 6 shows the $E$--$\omega$ plane for $E\in[m,m+\nu]$ and $\omega
\in[0,\nu]$ ($E$, however, cannot exceed its elastic endpoint
$E_\smallmax$). For a moment we will assume (as we did in drawing
fig.~6) that $\epsilon$ is smaller than
$\epsilon_\star=\nu'_\smallmin$, but we will later on free our
analysis from this constraint.  The vertical segments {\sf OP} and
{\sf PQ} indicate, respectively, the sets of points contributing to
the VS and HB corrections to the electron spectrum (see sect.~3) for a
specific value $E=E_0$ (in the virtual corrections it's simply
$\omega=0$).  The overall QED corrections to the electron spectrum are
obtained by adding up the VS and HB terms and clearly do not depend on
the value of $\epsilon$. In fig.~6, diagonal lines indicate sets of
points with the same value of the combined electron--photon energy.
The set of all points in the $E$--$\omega$ plane having the same
$\eom$ value consists of one or two segments, according to the
magnitude of $\eom$ which varies between $m$ and $m+\nu$. If $m \leq\eom
< m+\epsilon$, the photon energy is lower than the threshold
$\epsilon$ and the measured combined energy $\eom$ equals $E$. The QED
corrections to the differential cross section $d\sigma/dE_\omega$ will
coincide, in this case, with the VS corrections to the electron
spectrum $d\sigma/dE$ (see e.g.~segment {\sf AB}). If $m+\epsilon \leq\eom
\leq E_\smallmax$, the QED prediction for $d\sigma/dE_\omega$ will
consist of two contributions, namely the HB cross section
$[d\sigma/d(E+\omega)]_{\rm HB}$ (see e.g.~segment {\sf CD}) plus the
VS corrections to $d\sigma/dE$ (segment {\sf EF}).  Finally, if
$E_\smallmax <\eom < m+\nu$, $d\sigma/dE_\omega$ will be identical
to $[d\sigma/d(E+\omega)]_{\rm HB}$ (see e.g.~segment {\sf GH}). It is
important to notice that the complete QED prediction for the
differential cross section with respect to $\eom$ depends on the
threshold $\epsilon$, contrary to the complete QED prediction for the
electron spectrum computed in sect.~3.

We can summarize our results in a very simple way. The SM prediction
for the spectrum of the combined energy of electron and photon in
reaction (\ref{eq:nuegamma}) can be cast, up to corrections of
$O(\alpha)$, in the form
\bea
        \left[\frac{d\sigma}{dE_\omega}\right]_{\rm SM} \!\!\!\!
      &=& \!\!\frac{2mG_{\mu}^2}{\pi}
        \Biggl\{\gl^2(\eom) \!\left[\theta 
          +\frac{\alpha}{\pi} \flov(\eom,\epsilon,\nu) \right]
        +\gr^2(\eom) \left(1-\zom\right)^2 
        \!\left[\theta
        +\frac{\alpha}{\pi}\frov(\eom,\epsilon,\nu)\right]
                                \nonumber\\
      & & -\gl(\eom)\gr(\eom) \left(\frac{m \zom}{\nu}\right)
        \!\left[\theta+ \frac{\alpha}{\pi}
        \flrov(\eom,\epsilon,\nu)\right]\Biggr\},
\label{eq:SMdeom}
\eea
where $\zom=(\eom-m)/\nu$ and $\theta = \theta(E_\smallmax -
E_\omega)$.  As we mentioned in sect.~3, the deviations of the
functions $\gl(\eom)$ and $\gr(\eom)$ from the lowest--order values
$\gl$ and $\gr$ reflect the effect of the electroweak corrections (for
virtual corrections it is $\omega=0$ and $\eom=E$). The functions
$\fxov(\eom,\epsilon,\nu)$ ($X=L,R$ or $LR$), defined in the range $[m,
m+\nu]$, describe the QED effects discussed earlier in this section
(once again, for simplicity of notation, we will drop their $\nu$
dependence). Following our previous analysis, these functions
can be written in the very simple form
\be
     \fxov(\eom,\epsilon) = \fx^{\sss VS}(\eom,\epsilon) + 
                            \fxov^{\sss HB}(\eom,\epsilon),
\label{eq:fxovsum}
\ee 
where $\fx^{\sss VS}(\eom,\epsilon)$ are the ``exact'' VS corrections
of sect.~3 (eq.~(\ref{eq:vsexact})) and the functions $\fxov^{\sss
  HB}(\eom,\epsilon)$ are derived by dividing the $L$, $R$ and $LR$
parts of the above--studied HB cross section
$[d\sigma/d(E+\omega)]_{\rm HB}$ by $C\gl^2$, $C\gr^2(1-\zom)^2$ and
$-C\gl\gr(m \zom/\nu)$ respectively, with $C=2mG_{\mu}^2\alpha/\pi^2$.
The $\theta$ functions in eq.~(\ref{eq:SMdeom}) reflect the fact that
the lowest order prediction for $d\sigma/dE_\omega$ has a step at
$\eom = E_\smallmax$ and is zero if $\eom$ lies outside the elastic
range $[m, E_\smallmax]$.  The VS functions $\fx^{\sss
  VS}(\eom,\epsilon)$ are proportional to the same $\theta$ function,
while the corrections $\fxov^{\sss HB}(\eom,\epsilon)$ are set to zero
if $\eom \notin\; [m+\epsilon, m+\nu]$.

Earlier in this section we assumed $\epsilon < \epsilon_\star$
($\epsilon_\star=\nu'_\smallmin$).
Nevertheless, the $\epsilon_\star \leq \epsilon \leq \nu$ case can be
discussed analogously and our simple prescription (\ref{eq:fxovsum})
is valid in both cases.

In fig.~7 we show, as an example, the function $\flov(\eom,\epsilon)$
(thick) and its components $\flov^{\sss HB}(\eom,\epsilon)$ (medium)
and $\fl^{\sss VS}(\eom,\epsilon)$ (thin) for $\nu=0.862\mev$. The
solid (dashed) lines correspond to the choice $\epsilon=0.2\mev$
($\epsilon = 0.05\mev$).  Once again we would like to emphasize the
$\epsilon$ dependence of the complete QED corrections
$\fxov(\eom,\epsilon)$, to be contrasted with the $\epsilon$
independence of the $\fx(E)$ functions of sect.~3.

In figs.~8, 9 and 10 we compare the results of sects.~3 and 4. In
fig.~8 we chose $\nu=0.862\mev$ and plotted the functions $\fx(E)$
(thick) and $\fxov(\eom,\epsilon)$ for $\epsilon=$ 0.1 MeV (medium)
and 0.001 MeV (thin). In figs.~9 and 10 we plotted the same functions
with $\nu=10\mev$ ($\epsilon=$ 1 MeV, 0.1 MeV) and $\nu=1\gev$
($\epsilon=$ 100 MeV, 10 MeV), respectively.

We would like to remind the reader that the functions $\fx(E)$ can be
obtained from $\fxov(\eom,\epsilon)$ by simply setting $\epsilon=\nu$.
The limiting case $\epsilon=0$ was studied in detail in
ref.~\cite{DB2} (in particular, the results of the second article of
this reference were obtained, like ours, without employing the
ultrarelativistic approximation $E \gg m$).

%5555555555555555555555555555555555555555555555555555555555555555555555
\section{Discussion and Conclusions}

When are the results of sects.~2, 3 and 4 applicable? In sects.~2 and
3 we evaluated the $O(\alpha)$ SM prediction for the electron spectrum
in the reaction $\nu_l + e \rightarrow \nu_l + e \;(+\gamma)$
(eq.~(\ref{eq:SMdE})), where $(+\gamma)$ indicates the possible
emission of a photon. In this calculation we assumed that the
final--state photon is not detected and, as a consequence, we
integrated over all possible values of the photon energy $\omega$.
Therefore, eq.~(\ref{eq:SMdE}) is the appropriate theoretical
prediction to use in the analysis of $\nu$--$e$ scattering when the
detector is completely blind to photons of all energies, but can
precisely measure $E$, the energy of the electron. Of course, a
detector could provide more information by detecting photons as soon
as their energy is above an experimental threshold $\epsilon$.  In
this case, still assuming a precise determination of $E$, one can
employ eq.~(\ref{eq:SMdE}), minus its HB correction, to analyze those
events which are counted as nonradiative (elastic), while the HB part
can be used, at least in principle, for a separate determination of
the inelastic cross section. Indeed, contrary to previous
calculations, our predictions are valid for an arbitrary value of the
threshold $\epsilon$ (and include the previously unknown $LR$ term).

In sect.~4 we evaluated the spectrum of the total combined energy of
the recoil electron and a possible accompanying photon emitted in the
scattering process (eq.~(\ref{eq:SMdeom}); $\eom$ was defined in
eq.~(\ref{eq:eom})).  This type of analysis is useful when the photon
energy $\omega$ cannot be separately determined although it fully
contributes to the precise total energy measurement if its value is
above a specific threshold $\epsilon$. Let's consider an experimental
setup able to measure the photon energy if it's higher than
$\epsilon$, but completely blind to low energy photons $(\omega\!
<\!\epsilon)$.  Let's also assume that the electron energy $E$ is
precisely measurable independently of its value. This detector can
determine both the differential cross section $d\sigma/dE_\omega$
(eq.~(\ref{eq:SMdeom})) and the electron spectrum $d\sigma/dE$
(eq.~(\ref{eq:SMdE})) (as well as its separate HB component).  There
are experiments, however, which cannot measure $E$, but only $\eom$,
with a specific value of the threshold $\epsilon$. BOREXINO \cite{BX}
and KamLAND \cite{KL}, for example, are liquid scintillation detectors
in which photons and electrons induce practically the same response.
If a photon is emitted in the $\nu$--$e$ scattering process, its
energy $\omega$ is counted together with $E$, provided their sum lies
within a specific range. The appropriate theoretical prediction for
their analysis is given, therefore, by the cross section
$d\sigma/dE_\omega$ of eq.~(\ref{eq:SMdeom}) with a very small value
of $\epsilon$ (for the case $\nu=0.862\mev$ see the thin lines in
fig.~8). However, we should point out that although the QED
corrections $\fx(E)$ (in eq.~(\ref{eq:SMdE})) and
$\fxov(\eom,\epsilon)$ with small $\epsilon$ (in
eq.~(\ref{eq:SMdeom})) are different, their numerical values are very
small when $\nu=0.862\mev$, the energy of the monochromatic neutrinos
produced by electron capture on \Be7 in the solar interior.  In fact,
as shown in fig.~8, both $(\alpha/\pi)\fx(E)$ and
$(\alpha/\pi)\fxov(\eom,\epsilon)$ with small $\epsilon$ are in this
case of $O(\lsim 1\%)$, and neither of the above collaborations is
likely to reach this high level of accuracy in their analyses of the
crucial \Be7 line.

There are detectors in which it might not be possible to identify the
measured energy with either $E$ or $\eom$. Indeed, the electron and
the photon may produce indistinguishable signals and the total
observed energy might not be the simple sum of $E$ and $\omega$, but
some other function of these two variables. Super--Kamiokande
(SK)\cite{SK}, for example, a water Cherenkov counter measuring the
light emitted by electrons recoiling from neutrino scattering, uses
the number of hit photomultiplier tubes to determine the electron
energy. However, a photon emitted in the scattering process may induce
additional hits indistinguishable from those of the electron.
Moreover, a photon and an electron of the same energy may produce
different numbers of hits and, therefore, it might not be possible to
identify the total measured energy with the sum $E+\omega$.

SK measures solar neutrinos with energies varying from 5 to 18 MeV.
For $\nu=10\mev$, fig.~9 shows that the QED corrections to the
differential cross sections $d\sigma/dE$ (eq.~(\ref{eq:SMdE})) and
$d\sigma/dE_\omega$ (eq.~(\ref{eq:SMdeom})) are of $O(1\%)$.
Corrections of this order may be relevant for the analysis of the very
precise data obtained by this collaboration. In fact, SK's Monte Carlo
simulations of the expected energy spectrum of recoil electrons from
solar neutrino scattering include the QED corrections of
ref.~\cite{BKS} (as well as the EW ones). As we investigated in
sect.~3, these corrections provide good approximations of the complete
$O(\alpha)$ QED corrections $\fx(E)$ to the electron spectrum of
eq.~(\ref{eq:SMdE}) (see fig.~4). Our previous discussion, however,
seems to suggest that these corrections may not be appropriate for SK's
solar neutrino analysis.  On the other hand, the SM prediction for the
spectrum of the combined energy of electron and photon of sect.~4
(eq.~(\ref{eq:SMdeom})) may be suitable, but only if we can assume a
similar efficiency in the detection of photons and electrons, and if
also relatively low energy electrons contribute to the total energy
measurement.  If these conditions are not met, and the precision of
the data requires it, one should probably perform a dedicated analysis
of the double differential cross section $d^2\sigma/(dE \,d\omega)$
with a response function specifically designed for this detector. A
triple differential cross section $d^3\sigma/(dE \,d\omega \,d\phi)$,
where $\phi$ is the angle between the directions of the electron and
the photon, may also be useful (see the third article of 
ref.~\cite{ZKN}, and ref.~\cite{BBBS}).

We will conclude by noticing that the QED corrections, contrary to the
EW ones, depend strongly on the initial neutrino energy and become
sizeable at high energies. Their appropriate expression will have to
be taken into account in the analysis of future precise high energy
$\nu$--$e$ scattering experiments.

%%%%%%%%%%%%%%%%%%%%%%%%%%%%%%% ACKNOWLED %%%%%%%%%%%%%%%%%%%%%%%%%%%%%
\vspace{0.5cm}\begin{center}{\large\bf Acknowledgments}\end{center}

\noindent I would like to thank Prof.~J.~N.~Bahcall and
Prof.~A.~Sirlin for suggesting this problem and for enlightening
discussions. I am particularly indebted to Prof.~P.~Minkowski and
Prof.~A.~Sirlin for reading the manuscript and for very useful
comments. I would also wish to express my gratitude to J.~Arafune,
D.~Bardin, C.~Greub, A.~Held, K.~Holland, T.~Kinoshita, W.~Marciano,
M.~Nakahata, J.~Papavassiliou, M.~Samaras, M.~Schaden, S.~Schoenert, 
Y.~Takeuchi and Y.~Totsuka for instructive conversations and exchanges
of ideas. This work was supported by Schweizerischer Nationalfonds.

%%%%%%%%%%%%%%%%%%%%%%%%%%%%%%% REFERENCES %%%%%%%%%%%%%%%%%%%%%%%%%%%%
%\newpage

%%%%%%%%%%%%%%%%%%%%%%%%%%%%% FIGURES %%%%%%%%%%%%%%%%%%%%%%%%%%%%%%%%%

%%%%%%%%%% FIG.1 
\newpage
\begin{figure}[tbp]
\vspace{-2.8cm}\hspace{-3cm}\includegraphics[width=20cm]{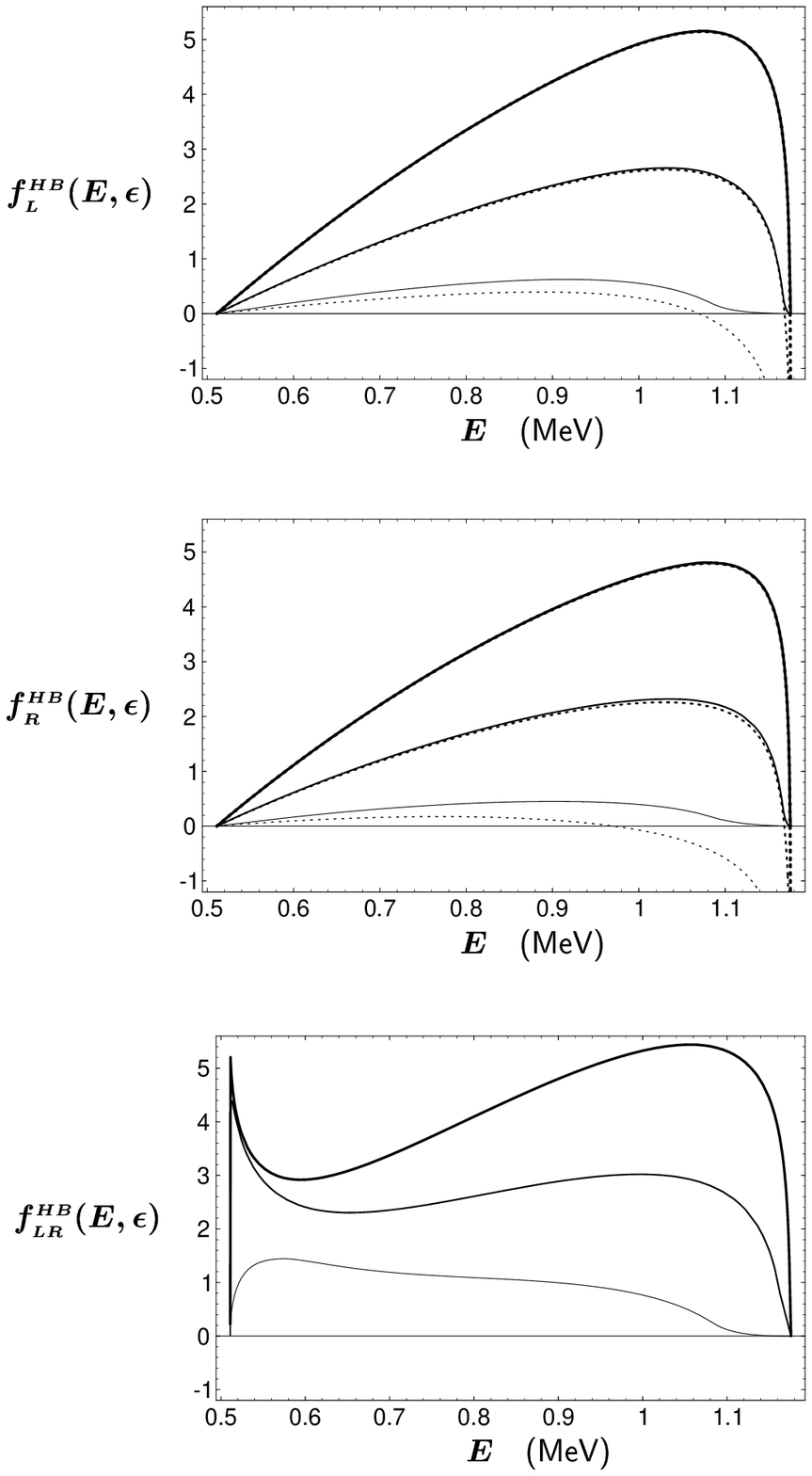}
\vspace{-7.5cm}
\caption{\sf Comparison of the \bm{$\fx^{\sss HB}(E,\epsilon)$}
  functions (solid lines) with their small \bm{$\epsilon$} analytic
  approximations (dotted lines) for an initial neutrino energy
  \bm{$\nu=$} 0.862 MeV. Thick, normal and thin lines indicate,
  respectively, \bm{$\epsilon=$}  0.001, 0.01 and 0.1 MeV.}
\label{figure:f1}
\end{figure}

%%%%%%%%%% FIG.2 
\newpage
\begin{figure}[tbp]
\vspace{-2.8cm}\hspace{-3cm}\includegraphics[width=20cm]{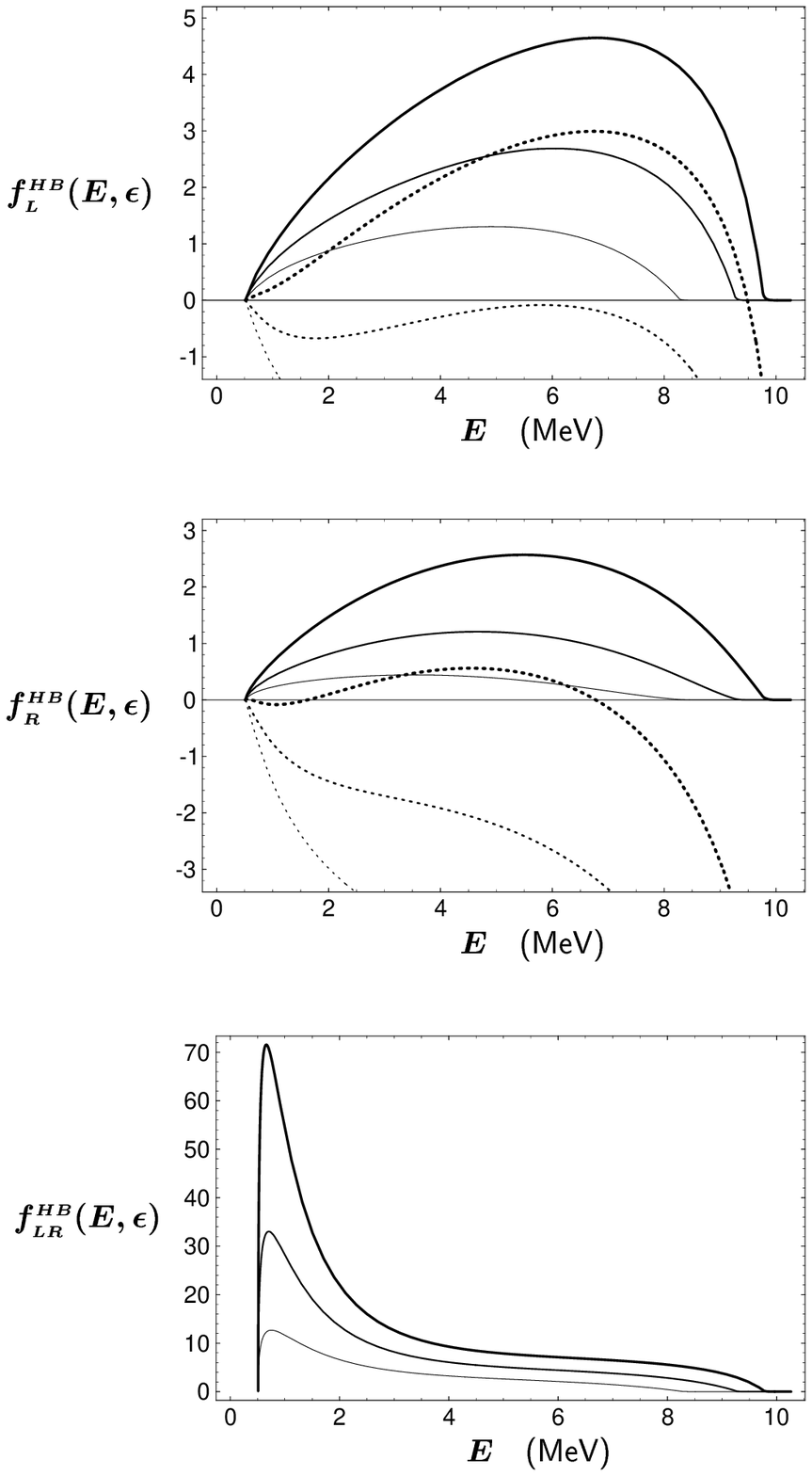}
\vspace{-7.5cm}
\caption{\sf Comparison of the \bm{$\fx^{\sss HB}(E,\epsilon)$}
  functions (solid lines) with their small \bm{$\epsilon$} analytic
  approximations (dotted lines) for an initial neutrino energy
  \bm{$\nu=$} 10 MeV. Thick, normal and thin lines indicate,
  respectively, \bm{$\epsilon=$}  0.5, 1 and 2 MeV.}
\label{figure:f2}
\end{figure}

%%%%%%%%%% FIG.3
\newpage
\begin{figure}[tbp]
\vspace{-2.8cm}\hspace{-3cm}\includegraphics[width=20cm]{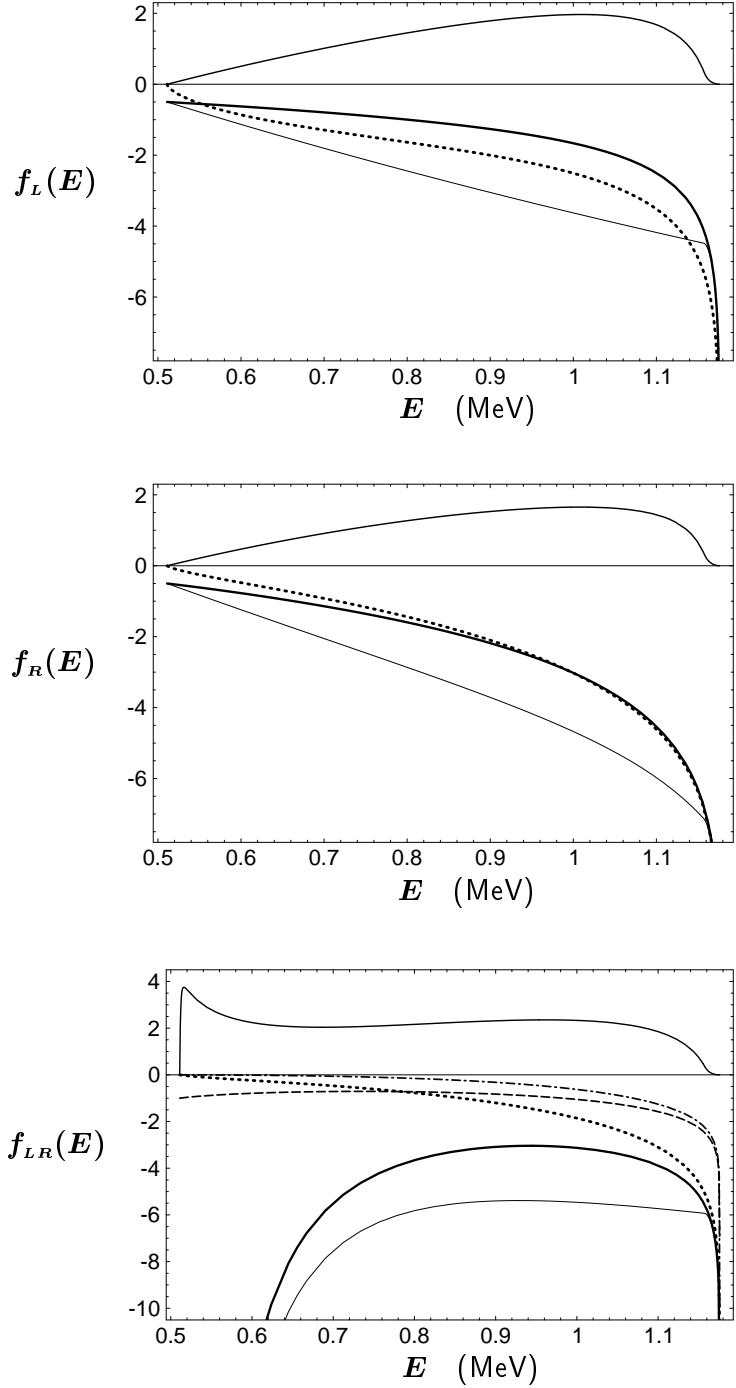}
\vspace{-7.5cm}
\caption{\sf 
  The functions \bm{$\fx(E)$} (thick solid), \bm{$\fx^{\sss
      HB}(E,\epsilon)$} (medium solid) and \bm{$\fx^{\sss
      VS}(E,\epsilon)$} (thin solid) for \bm{$\nu=$} 0.862 MeV and
  \bm{$\epsilon=$} 0.02 MeV. The dotted lines represent the
  \bm{$\fx(E)$} approximations of ref.~\cite{BKS}. In the $LR$ figure, 
  the dot-dashed line is the product of the \bm{$\flr(E)$}
  approximation of ref.~\cite{BKS} and \bm{$(mz/\nu)$},
  while the dashed line indicates the product \bm{$(mz/\nu)\flr(E)$}.}
\label{figure:f3}
\end{figure}

%%%%%%%%%% FIG.4
\newpage
\begin{figure}[tbp]
\vspace{-2.8cm}\hspace{-3cm}\includegraphics[width=20cm]{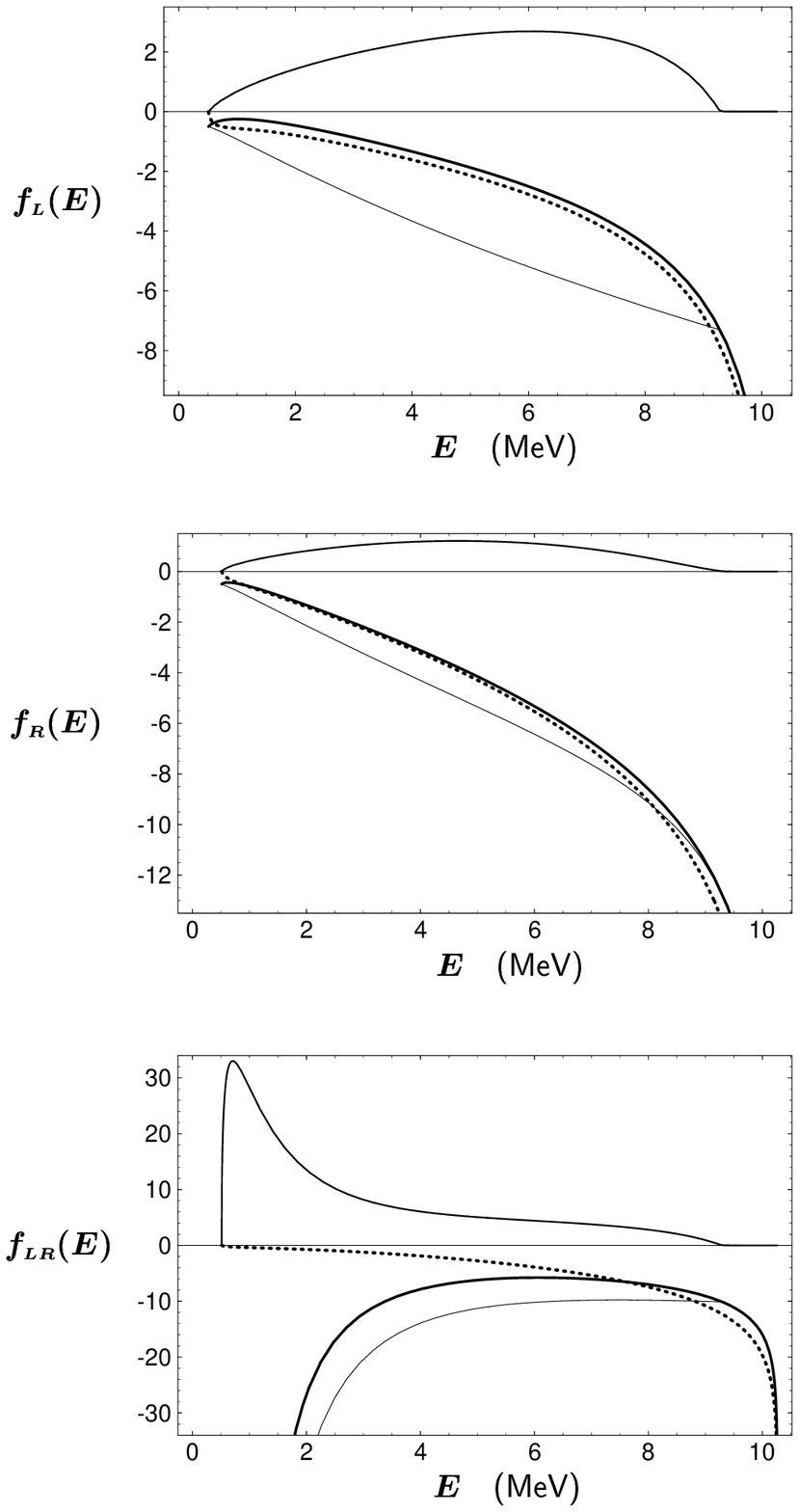}
\vspace{-7.5cm}
\caption{\sf 
  Same as Fig.~3, but for \bm{$\nu=$} 10 MeV and \bm{$\epsilon=$} 1 MeV.
  The dashed and dot-dashed lines are very close to zero and are not
  indicated.} 
\label{figure:f4} \end{figure}

%%%%%%%%%% FIG.5
\newpage
\begin{figure}[tbp]
\vspace{-2.8cm}\hspace{-3cm}\includegraphics[width=20cm]{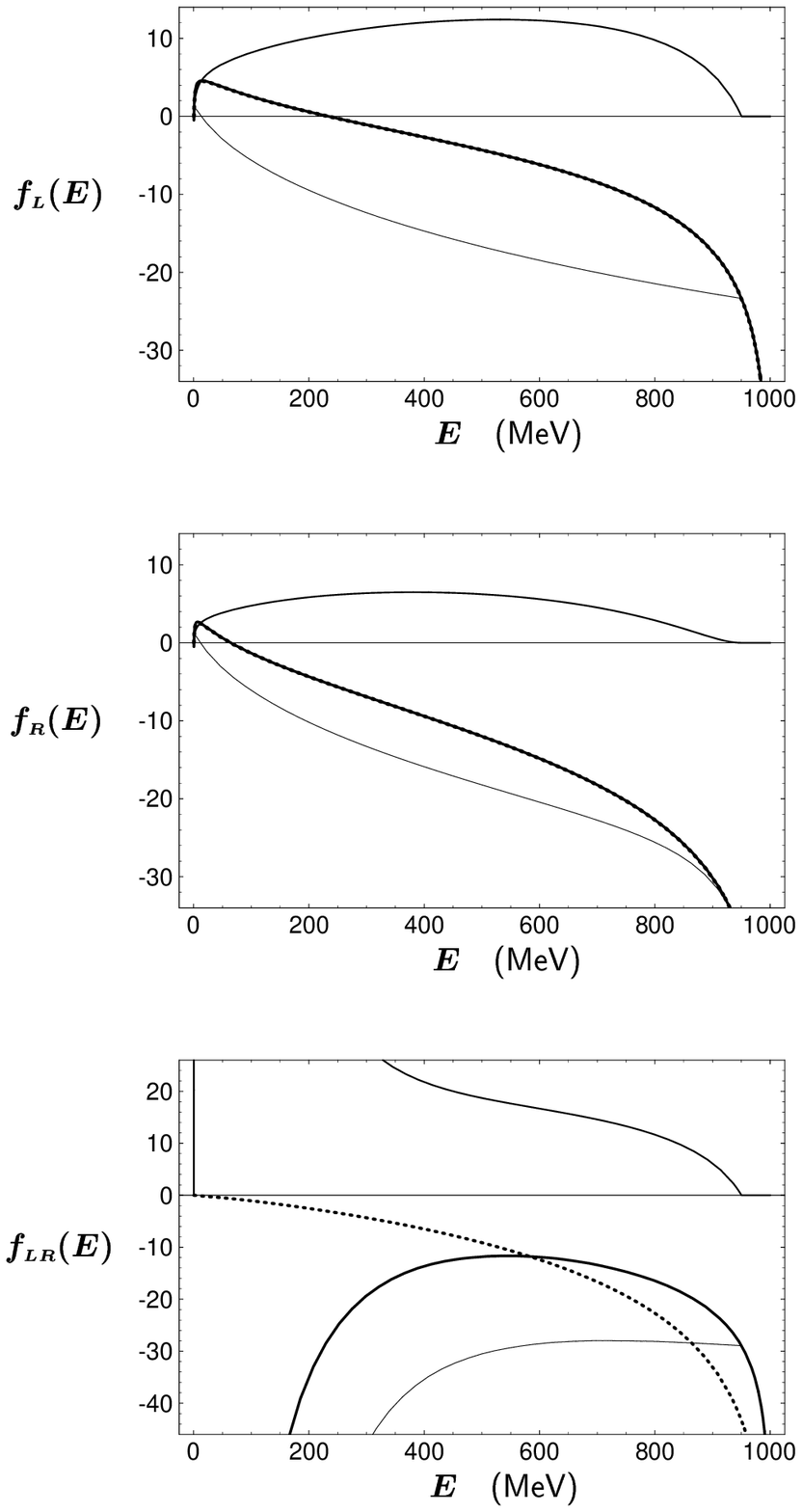}
\vspace{-7.5cm}
\caption{\sf 
  Same as Fig.~3, but for \bm{$\nu=$} 1 GeV and \bm{$\epsilon=$} 50 MeV.
  The dashed and dot-dashed lines are very close to zero and are not
  indicated.}
\label{figure:f5}
\end{figure}

%%%%%%%%%% FIG.6
\newpage
\begin{figure}[tbp]
\hspace{-2cm}\includegraphics[width=16cm]{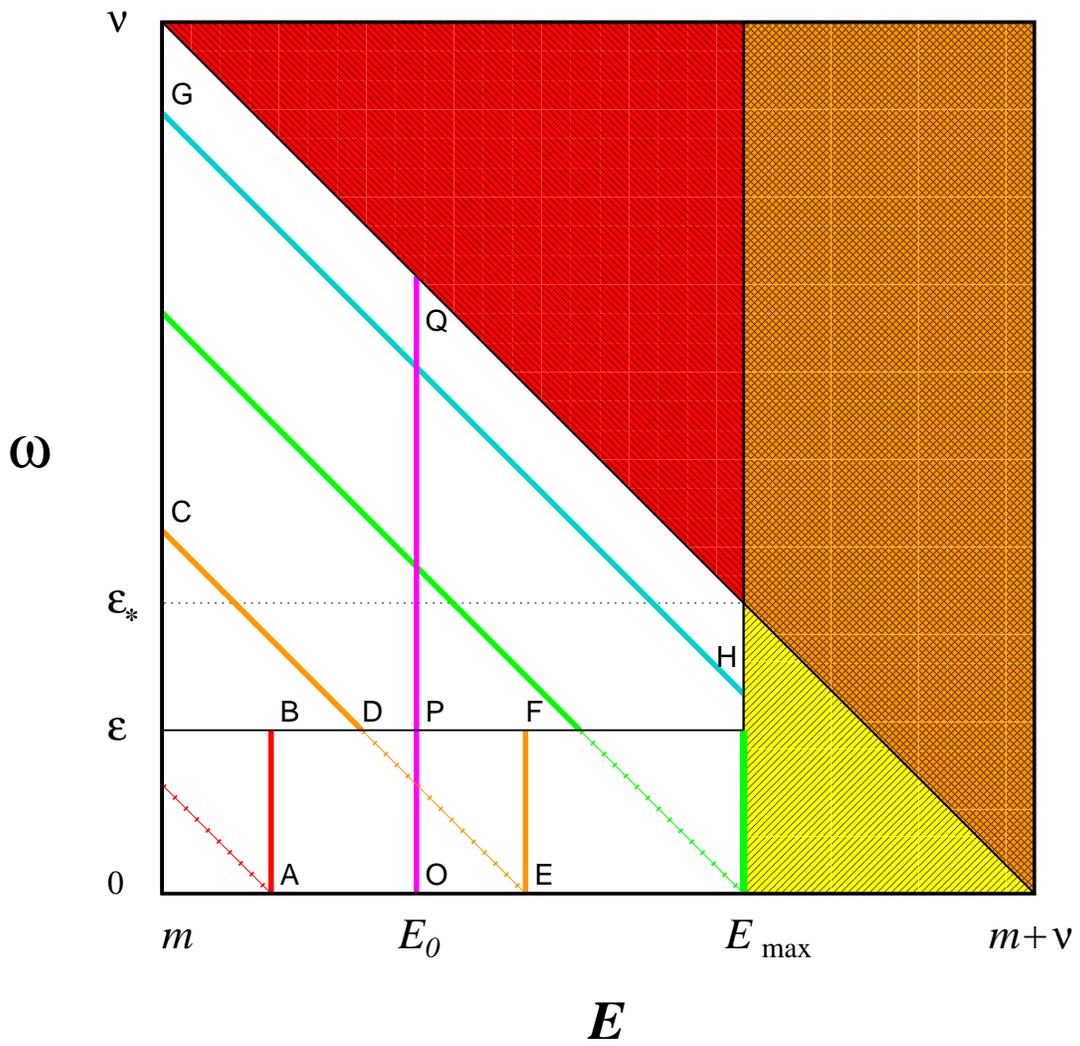}
\vspace{-4cm} 
\caption{\sf The \bm{$E$}--\bm{$\omega$} plane for
  \bm{$E\in[m,m+\nu]$} and \bm{$\omega\in[0,\nu]$}. See text 
  for details.}
\label{figure:f6}
\end{figure}

%%%%%%%%%% FIG.7
\newpage
\begin{figure}[tbp]
\vspace{-4cm}\hspace{-2cm}\includegraphics[width=20cm]{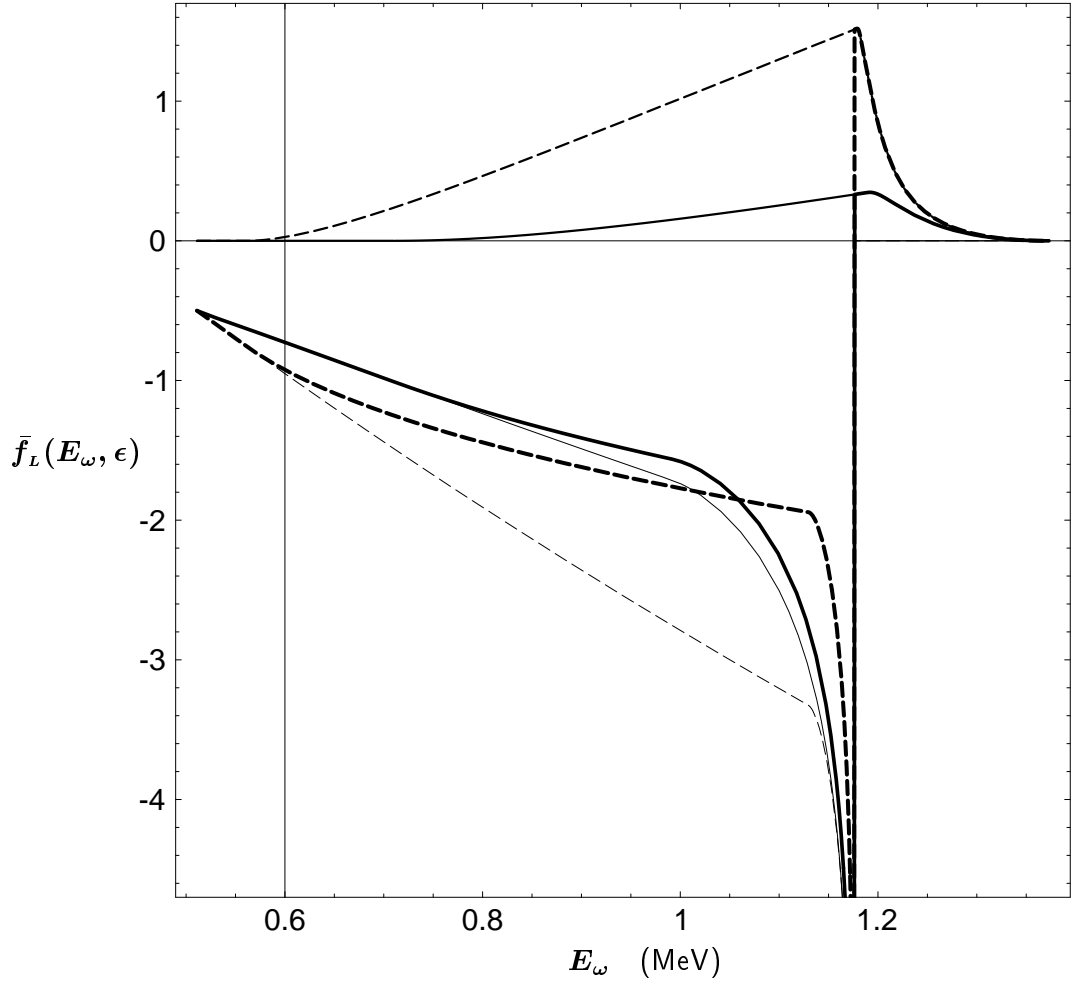}
\vspace{-8cm} 
\caption{\sf The function \bm{$\flov(\eom,\epsilon)$}
  (thick) and its components \bm{$\flov^{\sss HB}(\eom,\epsilon)$}
  (medium) and \bm{$\fl^{\sss VS}(\eom,\epsilon)$} (thin) for
  \bm{$\nu=$} 0.862 MeV. Solid and dashed lines indicate,
  respectively, \bm{$\epsilon=$} 0.2 and 0.05 MeV.}
\label{figure:f7}
\end{figure}

%%%%%%%%%% FIG.8
\newpage
\begin{figure}[tbp]
\vspace{-2.8cm}\hspace{-2.5cm}\includegraphics[width=20cm]{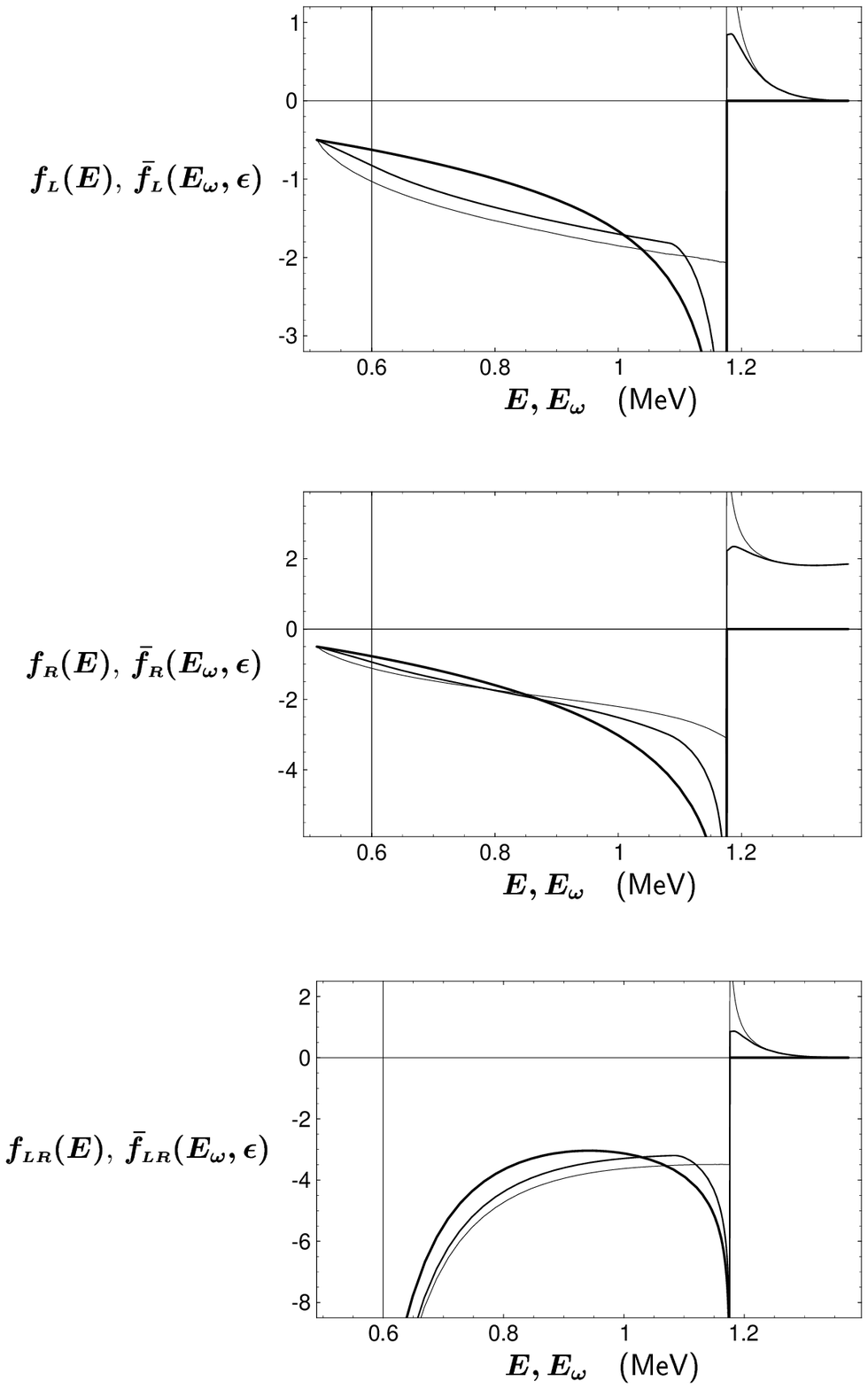}
\vspace{-7.5cm}
\caption{\sf The functions \bm{$\fx(E)$} (thick) and 
  \bm{$\fxov(\eom,\epsilon)$} for \bm{$\epsilon=$} 0.1 MeV (medium)
  and 0.001 MeV (thin). \bm{$\nu=$} 0.862 MeV.}
\label{figure:f8}
\end{figure}

%%%%%%%%%% FIG.9
\newpage
\begin{figure}[tbp]
\vspace{-2.8cm}\hspace{-2.5cm}\includegraphics[width=20cm]{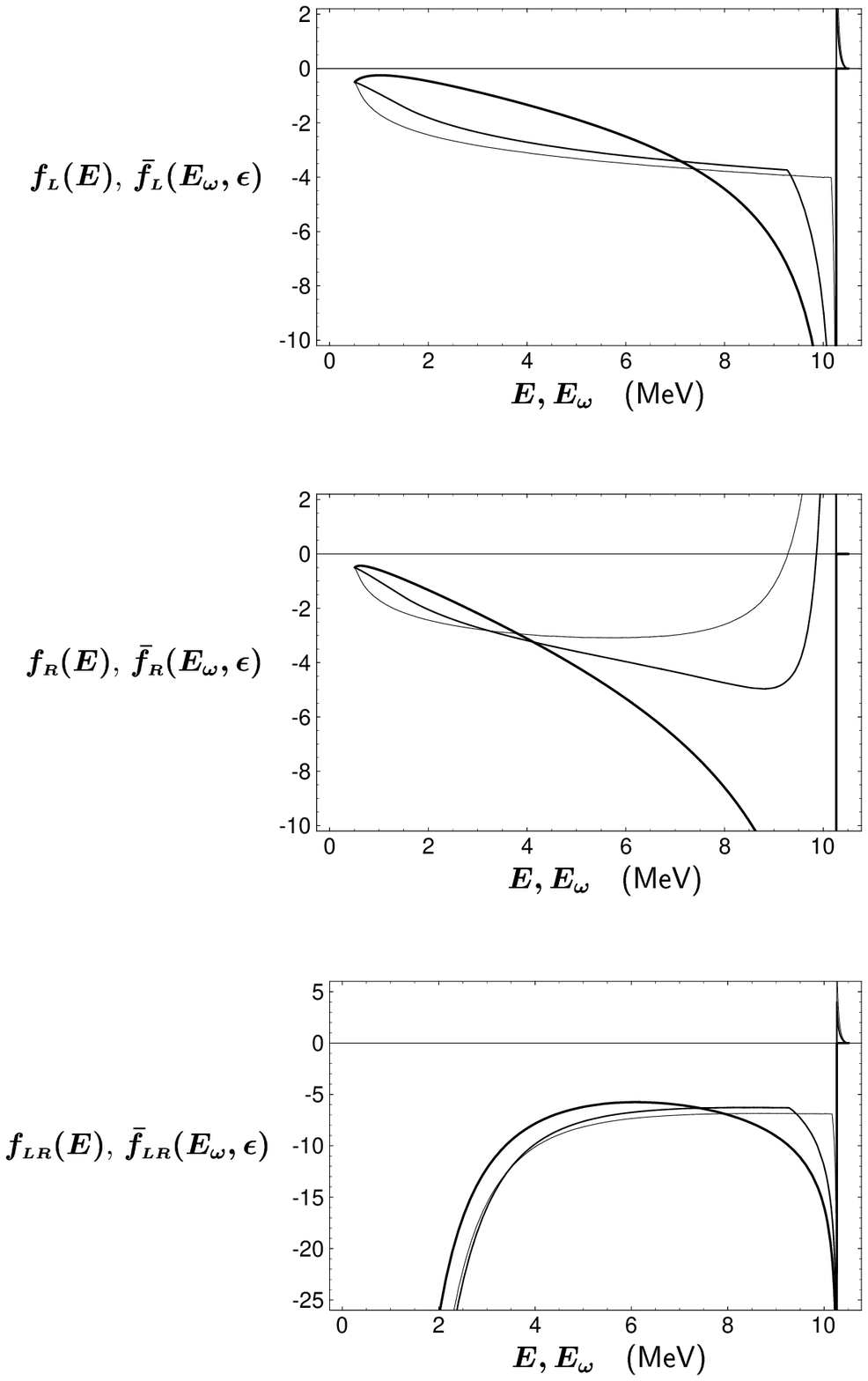}
\vspace{-7.5cm}
\caption{\sf \sf The functions \bm{$\fx(E)$} (thick) and 
  \bm{$\fxov(\eom,\epsilon)$} for \bm{$\epsilon=$} 1 MeV (medium) and
  0.1 MeV (thin). \bm{$\nu=$} 10 MeV.}
\label{figure:f9}
\end{figure}

%%%%%%%%%% FIG.10
\newpage
\begin{figure}[tbp]
\vspace{-2.8cm}\hspace{-2.5cm}\includegraphics[width=20cm]{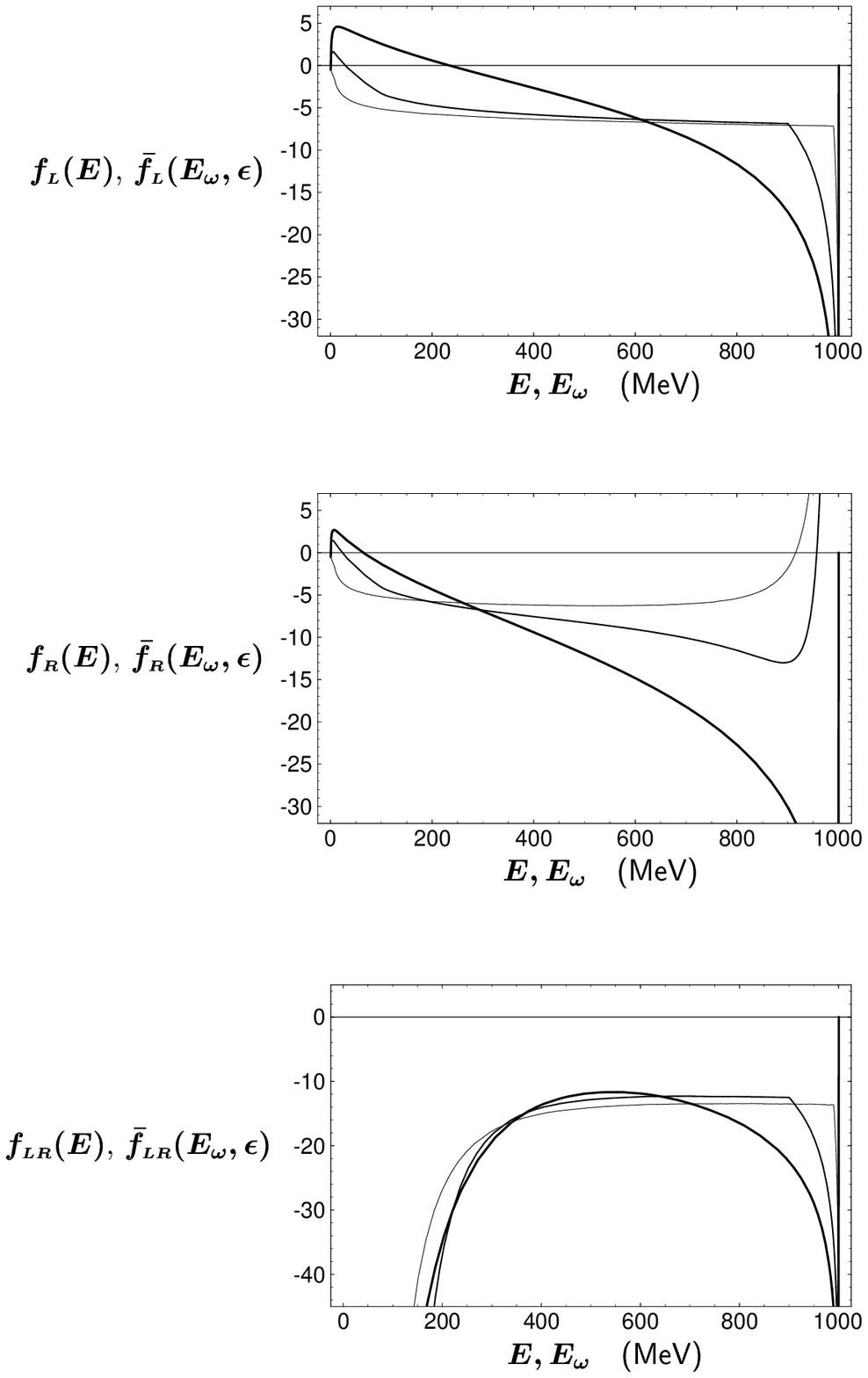}
\vspace{-7.5cm}
\caption{\sf \sf The functions \bm{$\fx(E)$} (thick) and 
  \bm{$\fxov(\eom,\epsilon)$} for \bm{$\epsilon=$} 100 MeV (medium)
  and 10 MeV (thin). \bm{$\nu=$} 1 GeV.}
\label{figure:f10}
\end{figure}

\end{document}